\documentclass[12pt]{article}

\usepackage{graphics,epsfig}
\usepackage{amsmath}

\begin{document}

\title{{\Large FASHION, NOVELTY AND OPTIMALITY:}\\
An application from Physics}
\author{Serge Galam$^1$\thanks{{\footnotesize galam@ccr.jussieu.fr}} and Annick
Vignes$^2$ \thanks{{\footnotesize annick.vignes@univ-reims.fr}} \\
$^{1}${\footnotesize Laboratoire des Milieux D\'{e}sordonn\'{e}s et
H\'{e}t\'{e}rog\`{e}nes, }\\
{\footnotesize \ University Paris 6, UMR 7603}\\
$^{2}${\footnotesize \ ERMES, University Paris 2 UMR 7017}$\ $}
\maketitle

\begin{abstract}
We apply a physical based model to describe the clothes
fashion market. Every time a new outlet appears on the market, it can invade
the market under certain specific conditions. Hence, the ``old'' outlet can
be completely dominated and disappears. Each creator competes for a finite
population of agents. Fashion phenomena are shown to result from a
collective phenomenon produced by local individual imitation effects. We
assume that, in each step of the imitation process, agents only interact
with a subset rather than with the whole set of agents. People are actually
more likely to influence (and be influenced by) their close ``neighbours''.
Accordingly we discuss which strategy is best fitted for new producers when
people are either simply organised into anonymous reference groups or when
they are organised in social groups hierarchically ordered. While
counterfeits are shown to reinforce the first strategy, creating social
leaders can permit to avoid them.
\end{abstract}



\newpage

\section{Setting the limits}

''Fashions have changed'', ''in fashion'', ''old-fashioned'' are phrases
frequently used in the street, on TV or in newspapers. But what fashion are
we talking about? The fashion of ideas, artistic fashion or, more
prosaically, dress fashion? When we talk about fashion, do we consider it to
be the result of a creative process (either intellectual or industrial), or
a method of communicating a certain ''way of life'' which would correspond
in economics to the level of information? In this paper, fashion should be
understood as the way a creative process entails new behaviour in a set of
consumers, independently of the information structure. We apply the model to
the clothes fashion market. Every time a new outlet appears on the market,
it can, under certain specific conditions, invade the market. Hence, the
''old'' outlet can be completely dominated and disappears. Each creator
competes for a finite population of agents.

As is the case for the movie industry or the industry of industrial design,
fashion activity depends on the level of creativity of the designers. It
also depends on the level of public recognition of this creativity. Nike
certainly shares this assumption, paying Michael Jordan (a famous basket
ball player in the U.S.) 20 million dollars in 1992, for endorsing Nike
running shoes.

Fashion does not improve either efficiency or the marginal utility of the
consumption. Clothes belong to a class of goods whose functional properties
are fundamental in their consumption. It is easy to understand why I need to
buy a dress, or a pair of jeans. It is harder to explain why I choose Adidas
shoes, or Levi's jeans. Moreover, it seems very difficult for an economist
to explain why women ask for short skirts one year, long skirts the year
after, and then short skirts again the year after that. Some theoricians may
talk of erratic preferences, but the desire to be in fashion could justify
Janssen and Jager's argument \cite{1} that ''...in satisfying their need for
identity, people may change their behaviour without changing their
preferences''.

In markets in general and the fashion market in particular, people are
interacting simultaneously, and constantly modifying their decisions. It is
very reminisence of the way atoms interact in inert matter. Galam and
Moscovici \cite{2} built a model of group decision making to describe the
dynamics of competing effect in updating individual opinions. De Oliveira et
al \cite{3} give some interesting applications of physic tools to study
economic and social problems has become more numerous. Social interactions
certainly play a major role in the understanding of fashion diffusion
process. By social interactions, ''we refer to the idea that the utility or
payoff an individual receives from a given action depends directly on the
choices of others in that individual's reference group, as opposed to the
sort of dependence which occurs through the intermediation of markets''
(Brock and Durlauf 1995). Hence, we postulate that a consumer maximises his
utility when he is dressed like everybody else in his peer group.

In this paper, we consider that fashion results from the desire both to
conform and to differentiate oneself from others. We also assume that
products are disembodied, thus any agent can switch from one brand to
another without cost. Under these assumptions, we attempt to explain how,
after a certain number of steps, a common unique behaviour can emerge from a
finite population with heterogeneous behaviour. We believe that the
structure and organisation of the market are strong factors in the diffusion
of fashion.

The fashion sector has changed since the 1950s. A lot of studies show that
this sector was organised into a very simple system until the 1970s
(creation, production, distribution). Since the beginning of the1980s, the
fashion sector has developed into a quite complex and destructured system,
with the emergence of different levels of quality (''Fashion off the peg''
and ''Luxury off the peg''). The fashion industry now follows a rationale of
''fili\`{e}re'', which implies a multiplication of brands (with a tough
policy of brand protection),and sales levels. This tough policy of brand
protection actually coexists with a high level of market piracy in the
design-based industry. For a long time the ''legal market'' fought hard
against this illegal market. It seems that things have recently changed: An
explanation could be in terms of of social welfare. Some recent analyses
show how if the counterfeit industry were to disappear it would leave the
place open to criminal organization, which would clearly be dangerous for
the economy and for society. According to us, allowing counterfeits can have
positive externalities for the main brands. A high level of piracy can be
interpreted as a signal that the product is ''fashionable''. In this sense,
the illegal market becomes an advertising tool. The more a brand is
counterfeited, the more this brand will be seen by a wide public.
Multiplying the levels of quality can also be interpreted as an advertising
tool. The consumer side can be seen as a lattice, in which people have
strong cliquishness with their peers, in hierarchic social groups. Hence,
the fashion market seems to be extremely hierarchical both on the demand and
on the offer side.

Proving the existence of a dominant strategy for the creators could help to
explain why some fashion goods emerge, where others fail. In what follows,
we assume that the aim of a new creator is to invade the fashion market. In
other terms, his product must become ''the fashion''. In this paper, we
compare the different strategies a brand can use to insure the widest
diffusion of its product with the minimum level of investment.We wish to
explain why the fashion ''fili\`{e}re'' is currently so diversified and
hierarchically organised, why counterfeits can have positive externalities
and why, therefore, creators must influence social interactions by
appropriate strategies and facilitate the diffusion of their creation.

\textbf{In part 2}, we focus on a market within which people are anonymous,
with homogeneous characteristics.We apply the concept and techniques of real
space renormalisation group from Physics, in the line of Ma \cite{4} to study
the fashion diffusion process within hierarchical structures. In particular,
we focus on the necessary conditions for a given brand to be sure to dress
the majority of people in a society.

In previous papers, Galam found that majority rule voting
produces a critical threshold to total power \cite{5,7}. The value of the critical
threshold to power is a function of the voting structure, namely the size of
voting groups and the number of hierarchical voting levels. We find here
that when producers consider consumers as simply being organised into
''reference groups''\footnote{{\footnotesize The ''reference group of an
individual possibly includes all the agents whom ''social psychologists''
call significant others''  \cite{8} . They can be friends, relatives,
neighbours, partners in business and so on.}} the presence of counterfeits
can be helpful in invading the market.

This result is demonstrated in \textbf{\ part 3} where we show that in the
case where consumers are anonymous, allowing counterfeits can be an
efficient tool to diminish the level of investment. \textbf{In part 4},
people are no longer anonymous. They are organised into small groups, which
can be interpreted as classified social groups with heterogeneous
characteristics. In this situation, people can recognise each other. We
postulate that they are \textit{a priori} leaders (they ''make'' the
fashion) or followers. Every time a follower imitates a leader, he becomes a
leader himself. We show that invasive creators can avoid counterfeits and
minimise their sunk costs if they can identify organised, hierarchic social
groups within the society.

\textbf{Part 5} deals with setting some quantitative conditions to invade a
given market with an already existing set of references groups and leaders. 
\textbf{Last part} concludes.

\section{From physics to fashion: The model}

When sufficient
agents in a consumer's social network switch from one product to another,
the preferences of this consumer are assumed to follow. But little is said
about how this socialisation appears, although some authors as McCauley,
Rozin and Schwartz  \cite {9} speak about cultural transmission. In what
follows, we show how creators can generate social effects by making useful
investments, then use these effects to invest less.

At the beginning of the game, all the consumers wear a specific brand, let
us say A. In the next period, a new creator B emerges and decides to invade
the market. To this end, he makes classical investments such as advertising
campaigns, promotions, etc. Now, people can choose between two brands (A and
B), representing two tendencies in the fashion. We assume that each
individual wants to be in fashion. Reference groups can be seen as cells
that will produce common social behaviour. The investment made by the new
creator will be efficient if it is sufficient to allow a wide diffusion
process.

At some stage, we consider all investments from both brands are stopped.
There, we denote by $p_{0}$ the overall proportion of people wearing the B
brand and $1-p_{0}$ the corresponding proportion wearing the A brand. From
these initial conditions we study the internal dynamics of spreading or
disappearence of the new brand B within the population. The underlying
assumption is that each individual wants to be in fashion. We believe that
it is not because they have erratic preferences that people will switch from
one fashion good to another, but rather because the maximisation of their
utility depends on their level of belongingness to a reference group.
Consuming the same product as their peers or friends increases this feeling
of belongingness.

To achieve this individual goal, we assume that each person goes through a
hierarchical imitation process. Each step of this process is related to a
reference social group that produces common social behaviour. Within this
framework, every reference group consists of the aggregation of other
smaller reference groups. Moreover, it is only when a local fashion is
established at one reference group level that the higher reference group
level becomes activated, as seen in Figure (1). Each increase in the
reference group size is referenced with a one time unit incrementation.

\begin{figure}[tbp]
\begin{center}
\centerline{\epsfysize=3.4cm\epsfbox{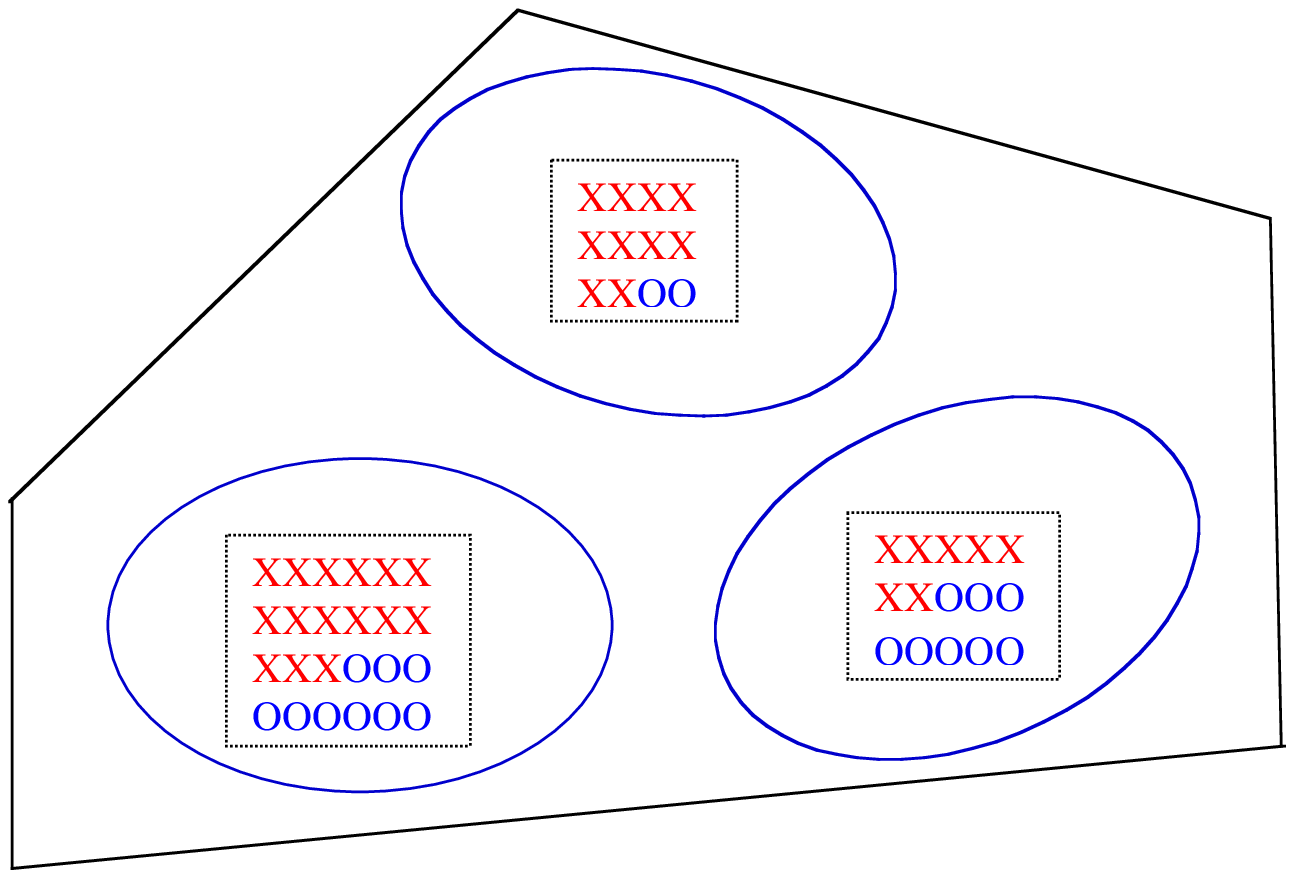}\epsfysize=3.4cm
\epsfbox{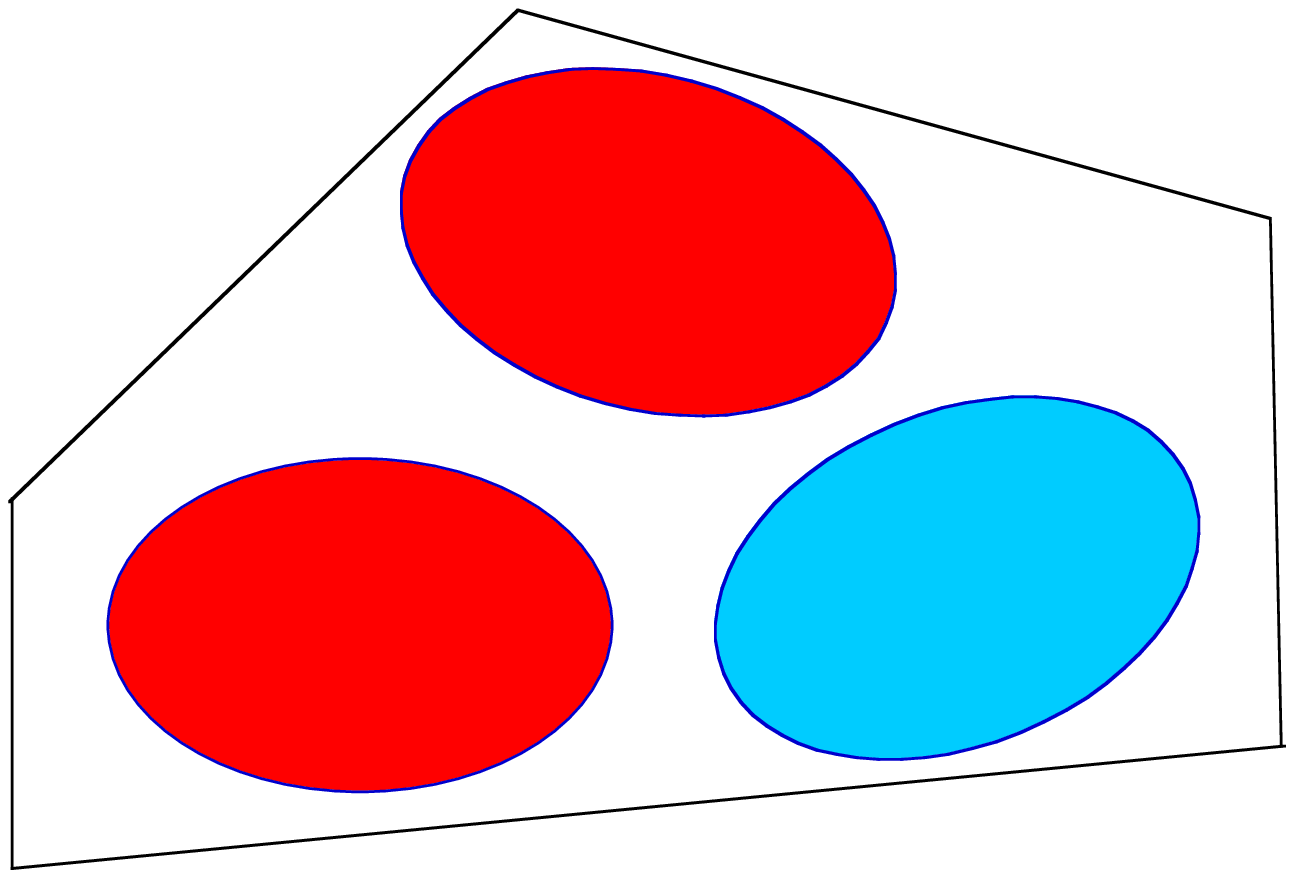}\epsfysize=3.4cm\epsfbox{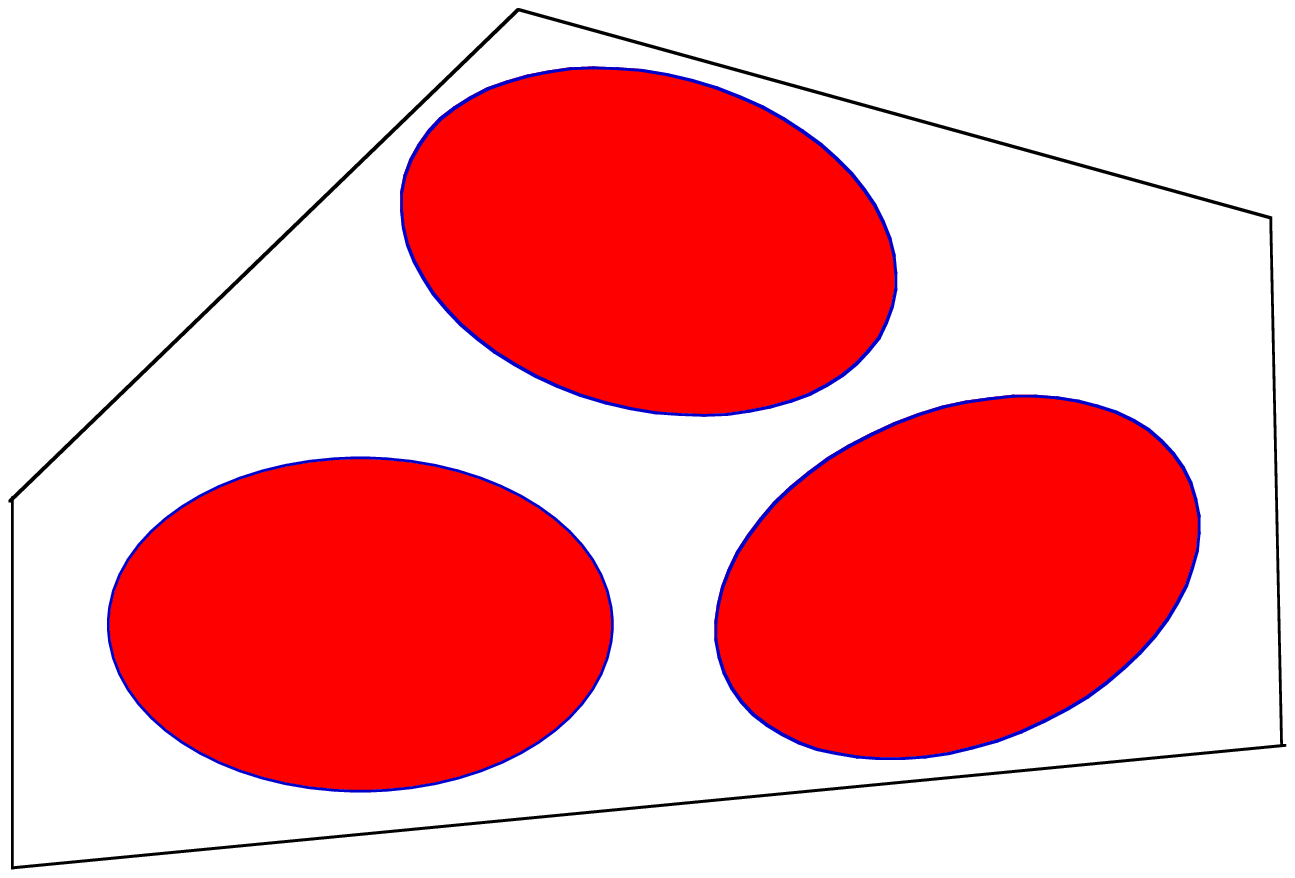}}
\end{center}
\caption{Left: First level of reference group with X and O behaviors. Common
behavior is not yet achieved. Middle: Common behavior has been completed at
first level reference group. X is associated to dark color and O behavior to
light color. Second reference group is activated. There common behavior not
yet completed. Right: Second reference group is activated. Common behavior
has been completed}
\end{figure}

\subsection{The anonymous in-fashion process}

We start from an anonymous situation in which the in-fashion model is driven
solely by a simple individual motivation to be like the majority of the
people around. There exists no leader, no intrinsic advantage to one brand
over the other, only the majority effect counts. We assume the population
dynamics of fashion awareness to be go within a successive set of reference
groups level extending to some factor r where r is an integer which counts
how many persons are within the first level of reference groups. Then to
simplify the equations we postulate the rescaling of the reference group
extension goes also by a factor r. It means people go in fashion first
according to a local majority among r persons. Once this step is completed,
the in-fashion process goes up to include r groups of r persons each, i.e.,
by majority among $r^{2}$ persons. And so on up to include the whole
population. For instance for $r=100$ we have 100 persons by first reference
group level and then it jumps to includes $r^{2}=10000$ and reach $%
r^{3}=1000000$, one million already after 3 levels.

To formalize above scheme we calculate the probabiliy $p_{1}$ to have one
initial reference group of r persons to have a majority of people wearing
the B brand starting from a whole larger population with a $p_{0}$
proportion of B wearing. All configurations of r persons having from r
persons wearing the B brand down to m ones where $m=\frac{r+1}{2}$ for odd r
and $m=\frac{r}{2}$ for even r add to yield a B majority. The case of
equality between the numbers of A and B wearing persons is attributed to the
B new brand as a tip to novelty. Accordingly having $p_{0}$ at $t=0$ leads
to $p_{1}$ at $t=1$ with, 
\begin{equation}
p_{1}=P_{r}(p_{0})\,
\end{equation}
where the function $P_{r}$ determines the renormalized proportion of B
wearing persons. More generally, starting from one reference level $n$ with
a B proportion $p_{n}$ leads to the new proportion $p_{n+1}$ at level $n+1$
with, 
\begin{equation}
p_{n+1}\equiv P_{r}(p_{n})=\sum_{l=r}^{l=n}\frac{r!}{l!(r-l)!}
p_{n}^{l}(1-p_{n})^{r-l}\ .
\end{equation}
Simultaneously the A proportion varies according to $1-p_{n+1}$.

\subsection{The optimal investment threshold: the necessary condition.}

\begin{quote}
For a creator who wishes to invade the market, an investment threshold
exists, below which investing is useless, no matter what the amount of
investment. Investing more is also superfluous. The optimal strategy
consists in setting the right level of investment just above the threshold
to enable the full market invasion driven by the in fashion process.
\end{quote}

To follow the dynamics of change in the density of \textit{B-dressed} people
we need to study the fashion function $P_{r}(p_{n})$ defined above in
Eq.(2). One shape is shown in Figure (2) for the case $r=101$. From the
Figure and Equation (2) it is found that the renormalized in-fashion process
produces a monotonic flow towards either one of the two stable fixed points $
p_{D}=0$ and $p_{I}=1$. The first one corresponds to the total disappearance
of the B brand with A preserving its initial monopoly. At the other extreme $
p_{I}$ represents a total B invasion with the A brand totally evicted.

\begin{figure}[tbp]
\begin{center}
\centerline{\epsfbox{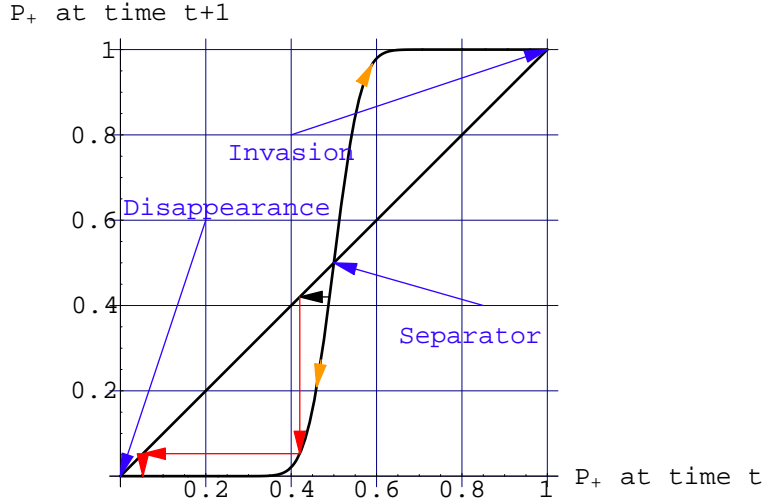}}
\end{center}
\caption{The fashion function $P_{r}(p_{n})$ for $r=101$ as function of $p$.
The separator is located at $50\%$ with the two attractors at $0$ and $1$.}
\end{figure}

In between these two points there exists another one $p_{c}$ which is
unstable since it produces the monotonic flow towards either one of the two
stable fixed points. It defines the critical density below which the
repeated extension in the in-fashion process leads inexorably to the total
disappearence of the B brand. For any odd reference group size it is located
at given by $p_{c}=\frac{1}{2}$ which gives the threshold to B invasion at
exactly $50\%$. Starting from $p_{0}<50\%$ leads towards 0 while the flow
leads to 1 for $p_{0}>50\%$. For instance in the case of $r=101$ we get the
following series starting from $p_{0}=0.40$, $p_{1}=0.02$ and $p_{2}\simeq 0$
In the case of $r=51$ we get the series is $p_{0}=0.40$, $p_{1}=0.07$ and $
p_{2}\simeq 0$.

Therefore the repeated in-fashion process produces the self-elimination of
any proportion of an initial B brand as long as $p_{0}<50\%$. To be
completed total disappearence only two reference levels are required.
Getting closer to the unstable fixed point increases slightly the number of
required reference levels. With $p_{0}=0.49$ the series are $p_{1}=0.42$, $
p_{2}=0.5$, $p_{3}\simeq 0$ and $p_{1}=0.44$, $p_{2}=0.21$, $p_{3}\simeq 0$
and for respectively $r=101$ and $r=51$.

At this stage, the key issue to ensure full monopoly for the initial
newcomer brand is a huge investment to guarantee a starting in fashion
process with more than fifty percent of the population wearing its B brand.
Any value less does make the all investment pure waste with the total
dispappearence of the brand. Such a condition put the level of success at an
almost impossible task.

\section{Counterfeits as a strategic tool}

\begin{quote}
When individuals on the consumer side are anonymous and simply organised
into undifferentiated large size reference groups, producers cannot
recognise them. In this situation, allowing counterfeits is a dominant
strategy, which will implicitly share the minimum level of investment needed
to invade the whole market.
\end{quote}

However, reaching a level below the threshold is just a waste of money. To
ignore the existence of a threshold level can yield to a quite expensive
strategy of penetration. On the other hand to pass the threshold is also a
waste of money. And moreover in the real world the exact value of the
treshold is unknow. This quite difficult situation in assessing the right
level of investment leads to view allowing counterfeits as a strategic
attitude.

It can indeed help in pushing the initial penetration above the critical
threshold, for no extra cost. When individuals on the consumer side are
anonymous and simply organised into undifferentiated reference groups,
producers cannot recognise them. In this situation, the best strategy is to
allow counterfeits, which will implicitly share the minimum level of
investment needed to invade the whole market. Therefore, allowing
counterfeits can be a useful strategy when producers are facing an anonymous
market, to avoid a total waste of the initial investment.

For instance, considering an investment yielding some initial $p_{0}<p_{c}$
leads as seen before to zero. At constrast allowing counterfeits to yield
some $q_{0}$ of persons wearing the counterfeits produces an effective
proportion of B wearing $p_{0}+q_{0}$ such that now $p_{0}+q_{0}>p_{c}$. In
that case the B will invade the market thanks to the counterfeits. Altough
it does hold the one hundred percent of the full market it does have make a
profitable investment.

In this sense, counterfeits can be considered as a diffusion vector and
then, be a useful tool for lowering the threshold.

But this tool can also have dangerous effect. For example, the counterfeit
producer can invade the market, creators never know exactly until where the
counterfeits will invest, the presence of a high number of counterfeits can
make the brand less attractive for a certain part of the population. We will
now explore whether creators can use other tools to decrease the level of
initial investment.

\section{Imitating your neighbours: smaller size groups}

We have already shown that invading a market can have a high cost, or/ and
be an inefficient strategy when the threshold is not reached. In this case,
creators may be tempted to allow counterfeits in order to decrease their
level of investment. However such a strategy does have also a high cost. On
this basis we suggest another strategic tool consisting in acting on the
in-fashion frame itself.

The idea is to modify the value of the unstable threshold which is
instrumental in determining the all or nothing investment to invade or
disappear from a market. To shift the threshold from the $50\%$ value
towards a lower value for the invader means to increase it to the in market
brand. Splitting the unstable threshold value into two different values
implies to introduce some asymmetry to favor the new coming brand. This can
be done naturaly since we already have introduced a bias in favor of the new
brand. Indeed we assume that given a reference group, in case of a tie,
equal A and B wearing, the in fashion process leads to choose the new brand
as the tip for novelty.

However for large reference group sizes, the occurrence of a tie is very
rare and does not have much effect on the overall dynamics. At constrast
smaller reference groups will exhibit more often tie situations. Therefore
to implement our new strategy requires to decrease the reference size
groups. Such a change can be obtained by producing clear social signs to
make people able to recognize each other. We are suggesting taking out from
anonymity the imitating processes to turn it personalization. In others
words make people choose whom they wish to resemble. To make our task
easier, we talk about ''neighbours''.

\subsection{From anonymity to neighbourhood}

\begin{quote}
For a creator who wishes to invade the market but avoid counterfeits, an
optimal strategy consists in breaking down the size of reference groups by a
personalization of the in fashion process. The smaller the reference groups,
the lower the investment level.
\end{quote}

Having find out where to act to produce a disymmetry in favor of the
incoming new brand, we now analyse our model in the case of maximum tie
effect which occurs at the smaller social group exhibiting a tie, a group of
4 persons. If at time $t=0$, 3 or 4 people in the group are \textit{
A-dressed },then all the $4$ people end up \textit{A-dressed} at time $
t=0+1=1$. However if only 0 or 1 people are \textit{A-dressed}, then all $4$
people end up \textit{B-dressed}. In case of a tie with 2 \textit{A-dressed}
and 2 \textit{B-dressed}, a bias in favour of novelty results in the new
brand B being adopted by the whole group. Accordingly putting $r=4$ in Eq.
(2) gives for the first reference group level in fashion result, 
\begin{equation}
p_{1}=P_{4}(p_{0})=p_{0}^{4}+4p_{0}^{3}(1-p_{0})+6p_{0}^{2}(1-p_{0})^{2}\ ,
\end{equation}
at time $t=1$, where as before $p_{0}$ is the initial proportion of B
brand-wearing people at $t=0$. The following three configurations \{(2 A, 2
B), (4 A, 3B), (0 A, 4 B)\} leads to (0 A, 4 B). While the in fashion
function $P_{4}(p_{n})$ still have the two stable fixed points zero and one,
the unstable fixed point is now located at, 
\begin{equation}
p_{c}=\frac{5-\sqrt{13}}{6}\simeq 0.23\ ,
\end{equation}
which put the threshold to B invasion at about $23\%$, a much lower value
than $50\%$ as shown in Figure (3).

\begin{figure}[tbp]
\begin{center}
\centerline{\epsfbox{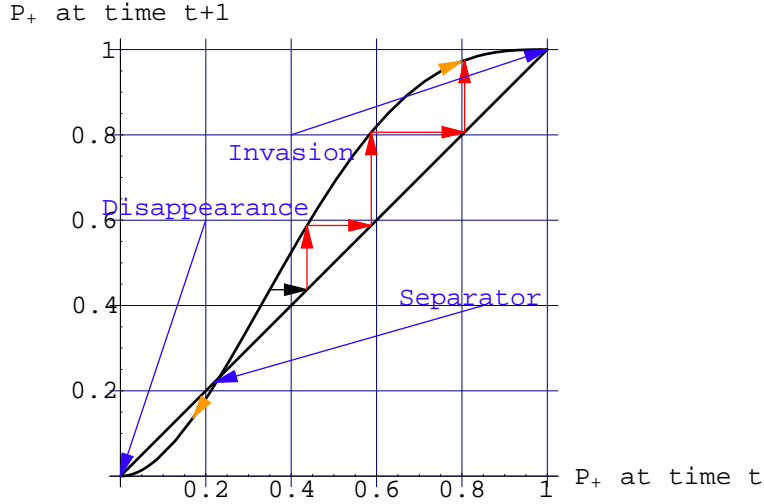}}
\end{center}
\caption{The fashion function $P_{r}(p_{n})$ for $r=4$ as function of $p$.
The unstable fixed point is located at $23\%$.}
\end{figure}

At the same time to keep on the whole market the A brand must always share
more than $77\%$ of the market. Starting from $p_{0}<23\%$ leads towards 0
while the flow leads to 1 for $p_{0}>23\%$.

The above people, all locally (groups of four) wearing the same brand,
constitute the first level of the hierarchy reference groups denoted level
1. Then the same in-fashion process is repeated, but now within reference
groups of 16 people each. Again, each whole reference group of 16 people
adopts the same brand, either A or B, according to a local majority rule. A
tie situation still yields a B brand choice. In terms of density the same
equation holds, with 
\begin{equation}
p_{2}=p_{1}^{4}+4p_{1}^{3}(1-p_{1})+6p_{1}^{2}(1-p_{1})^{2}\ ,
\end{equation}
since with the three types of configurations (8 A, 8 B), (4 A, 12 B), (0 A,
16 B) which can be noted as follows: 4\{(2 A, 2 B), (1 A, 3 B), (0 A, 4 B)\}.

Each additional increment of time increases the size of the reference group
with a new in-fashion process leading to an homogenisation on brand A or B
with the density at level $n+1$ given by 
\begin{equation}
p_{n+1}\equiv
P_{4}(p_{n})=p_{n}^{4}+4p_{n}^{3}(1-p_{n})+6p_{n}^{2}(1-p_{n})^{2}\ ,
\end{equation}
where $p_{n}$ is the proportion of B brand-wearing people at level n with
the following three configurations $4^{n}$\{(2 A, 2 B), (1 A, 3B), (0 A, 4
B)\}.

To illustrate the quantitative changes produced by the smaller size effect,
we first come back to the case of an intitial $p_{0}=0.40$ which goes down
to zero within two reference group levels for $r=100$ and $r=51$. Now in the
case $r=4$ not only the series does not go to zero but instead increases
toward one, i.e., the full market invasion with $p_{1}=0.53$, $p_{2}=0.72$, $
p_{3}=0.93$ and $p_{4}=1$. Only 4 reference groups levels are enough to
total invasion. At the fourth level only 96 people are involved instead of
the one million after 3 levels in the case $r=100$.

But now, as the threshold is at $23\%$ we can follow what happens with an
initial investment accounting for instance to a B wearing proportion $
p_{0}=0.25$. we have the series $p_{1}=0.26$, $p_{2}=0.28$, $p_{3}=0.32$, $
p_{4}=0.38$, $p_{5}=0.48$, $p_{6}=0.66$, $p_{7}=0.89$,and $p_{8}=1$. Eight
levels of reference groups are required for a total invasion of the market.
An initial value $p_{0}=0.30$ yields $p_{1}=0.35$, $p_{2}=0.43 $, $
p_{3}=0.58 $, $p_{4}=0.80$, $p_{5}=0.97$ and $p_{6}=1$ reducing the number
of reference groups required to six.

One illustration of the different outcome resulting from breaking down the
neighborhood is shown in Figures (4, 5, 6, 7) for a total population of 64
persons with 21 dressing an X item and 43 a O item, X being the new one.
Figure (4) shows the in-fashion process at work for one reference group
embodying at once the whole population of the 64 persons. The result is the
winning of the item O initially chosen by the majority.

Figures (5, 6, 7) shows the same initial population with now the existence
of 3 discrete reference groups with respectively 4, 16 and 64 people. Figure
(5) shows the introduction of the first level of neighboring reference group
with four persons each. X and O wearing are present on the left side. Common
behavior is completed on the right side with now 24 X and 60 O.

In Figure (6) new larger reference groups of 16 people each are activated as
shown on the left side. On the right side common behavior has been completed
with an equality between X and O, each item being chosen by 32 persons..
Figure (7) exhibits the last reference group which embodies the whole
population as shown on the left side. On the right side everyone is wearing
an X item. The opposite outcome of Figure (5 showing how the creation of
intermediate reference group levels has been able to reverse the outcome of
the in-fashion process.

\begin{figure}[t]
\begin{center}
\centerline{\epsfysize=7cm\epsfbox{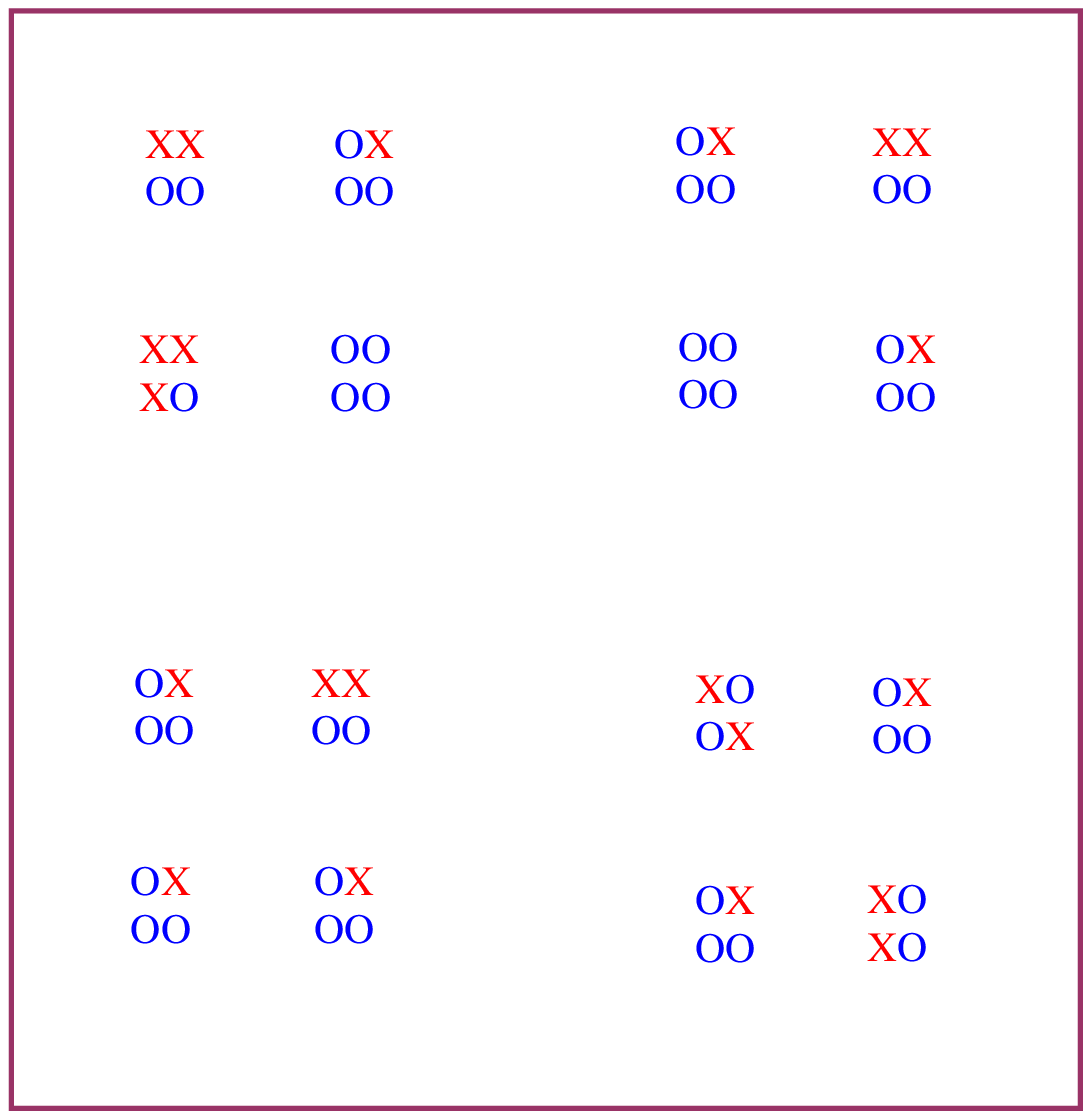}
\epsfysize=7cm\epsfbox{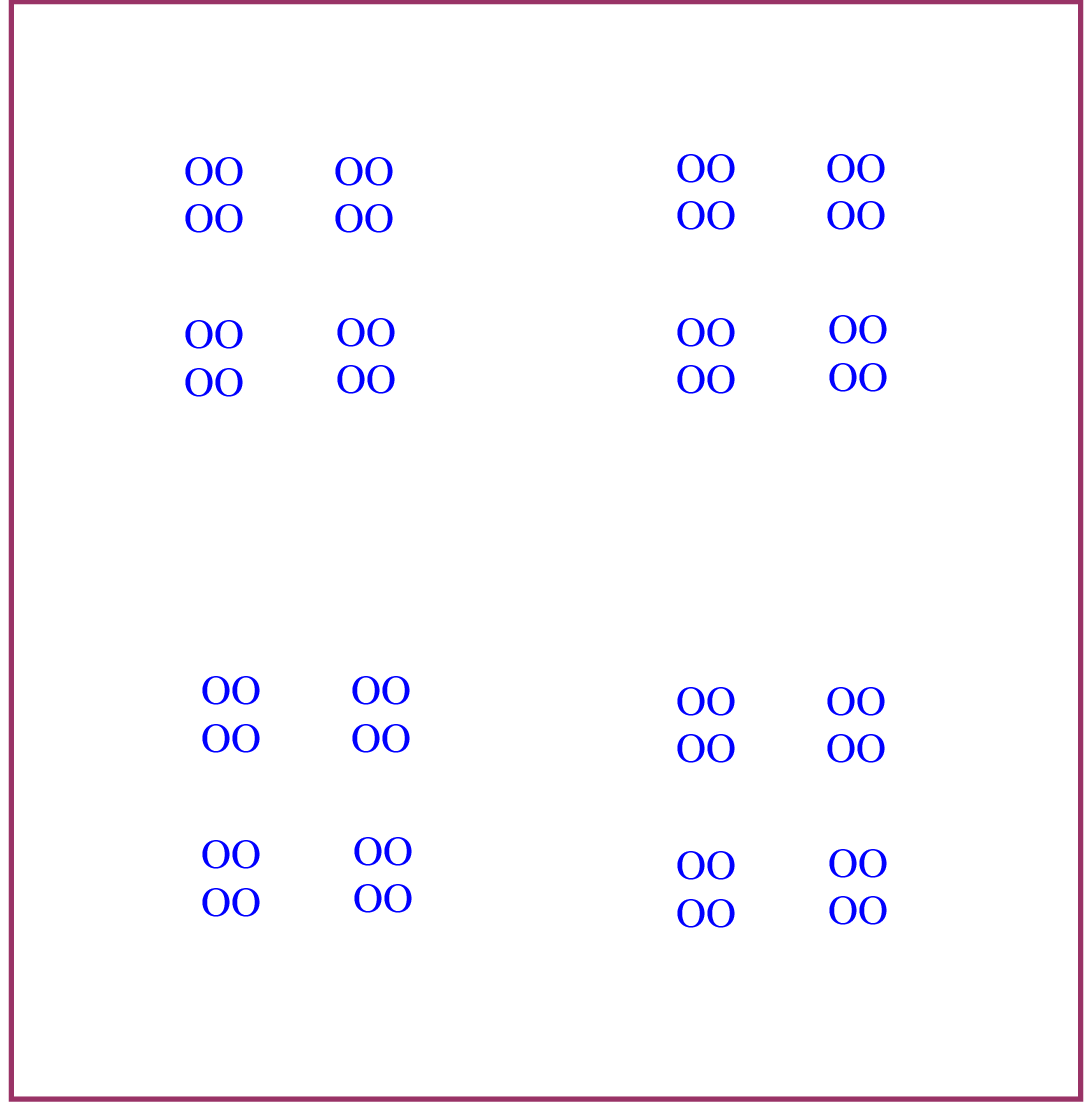}}
\end{center}
\caption{One group of 64 persons whom 21 are X wearing and 43 are O wearing
on the left side. Common behavior is not yet achieved . The in-fashion
process has been completed on the right side where everyone is wearing the O
item shared by the initial majority.}
\end{figure}

\begin{figure}[t]
\begin{center}
\centerline{\epsfysize=7cm\epsfbox{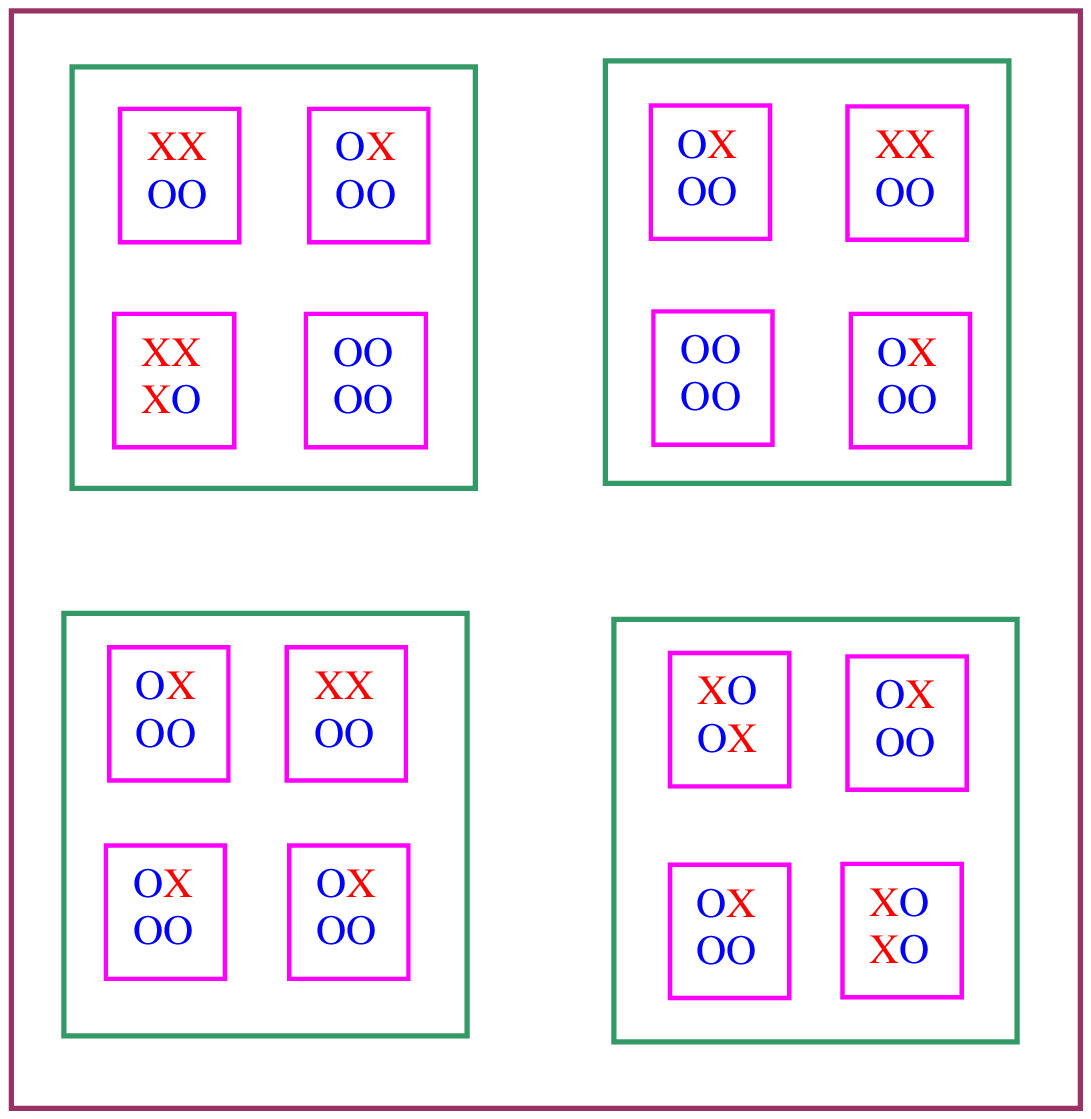}\epsfysize=7cm
\epsfbox{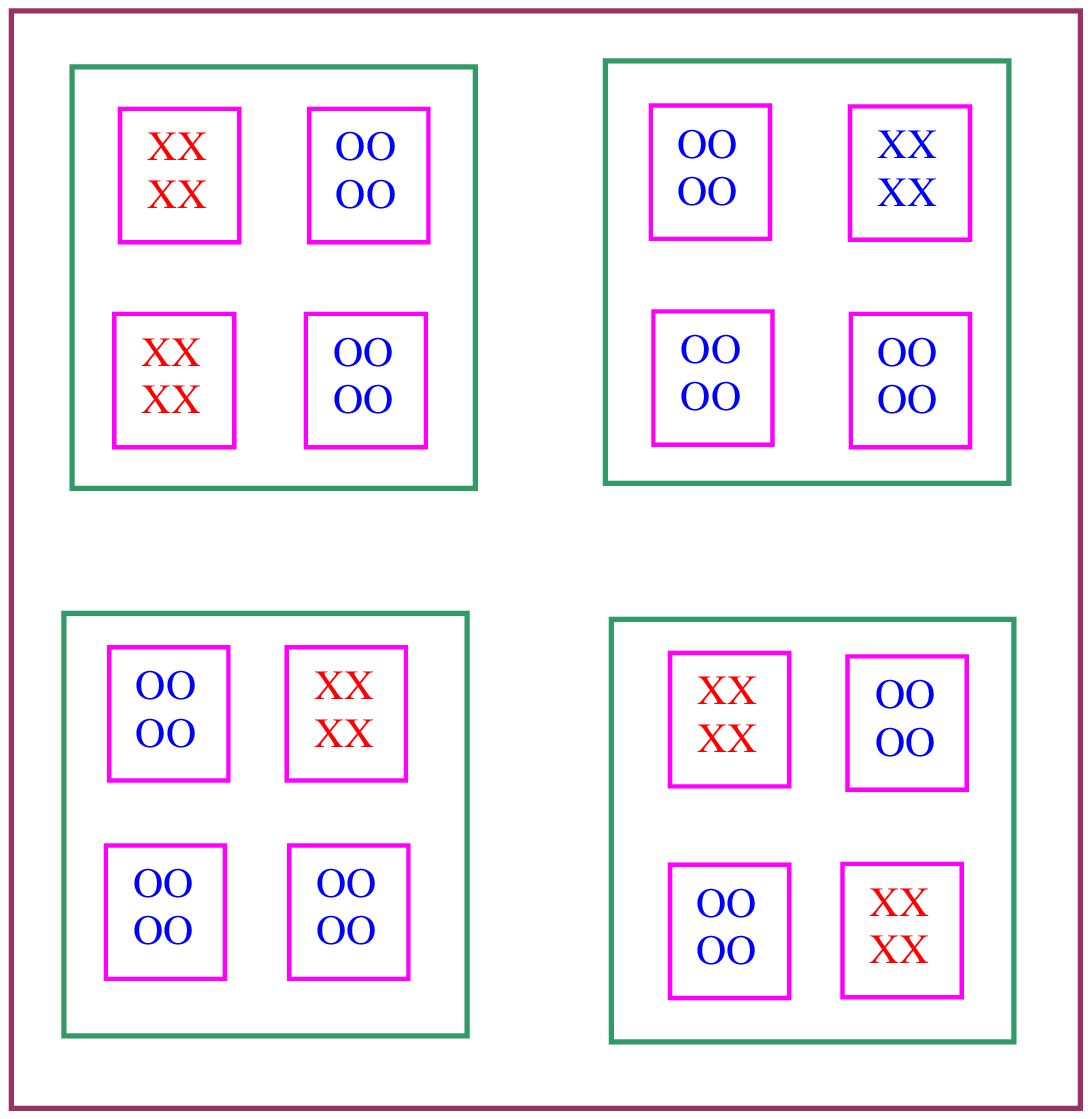}}
\end{center}
\caption{Left: First level of reference group with four persons each. X and
O wearing are present. Right: Common behavior has been completed at first
level reference group with 24 X and 40 O. }
\end{figure}

\begin{figure}[t]
\begin{center}
\centerline{\epsfysize=7cm\epsfbox{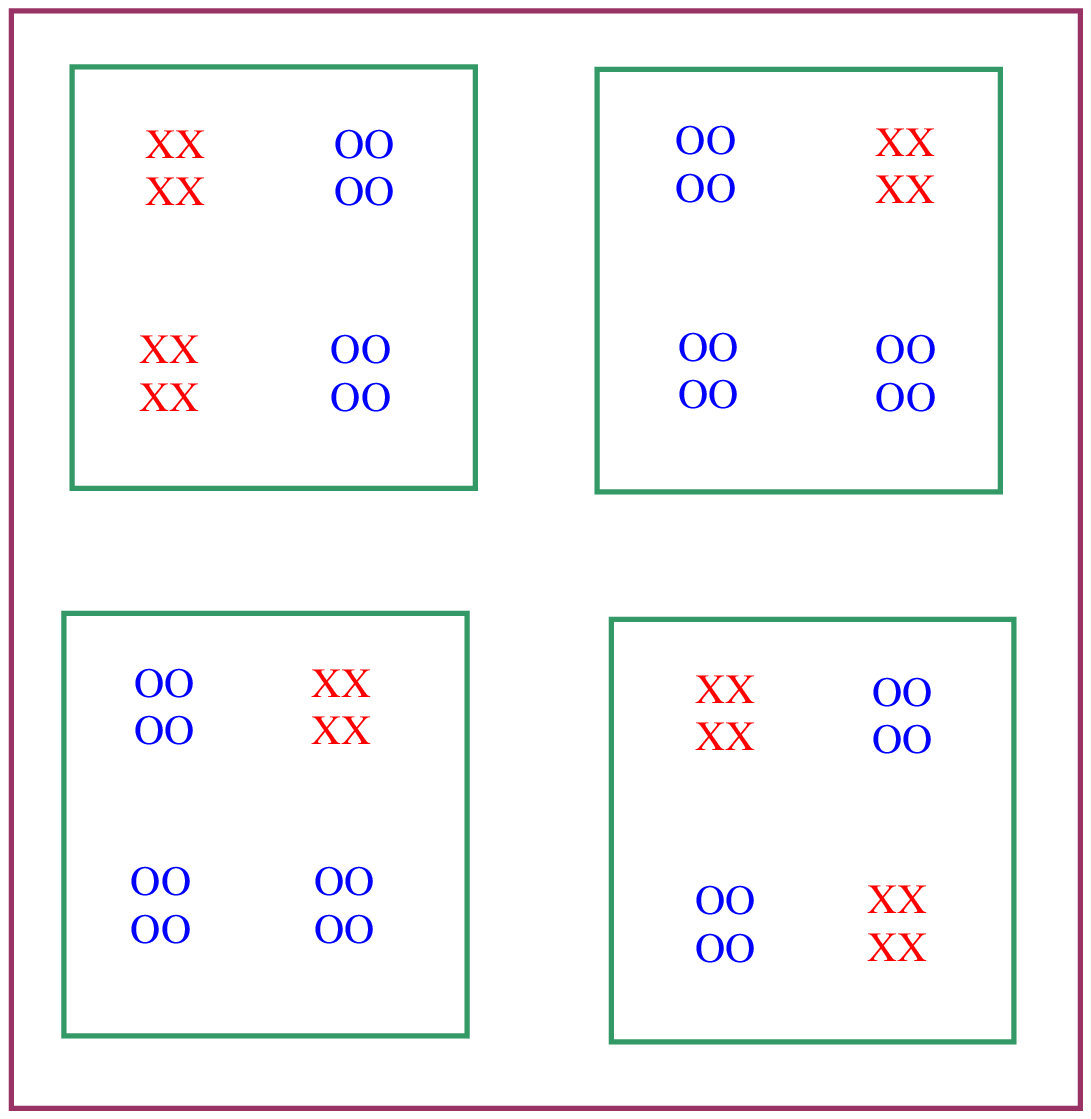}
\epsfysize=7cm\epsfbox{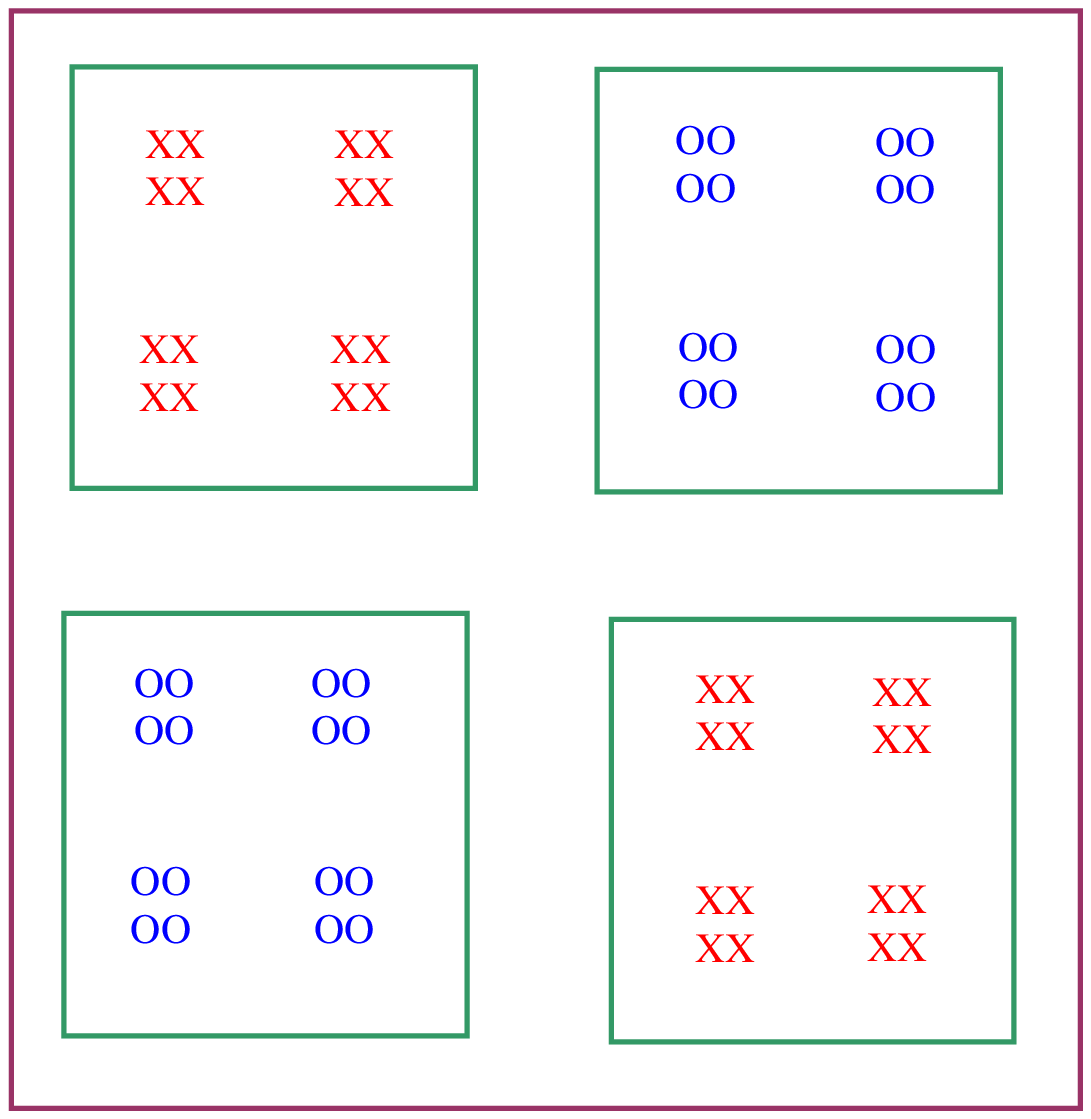}}
\end{center}
\caption{Left: Left: Second level reference group is activated to include
groups of 16 persons each. Common behavior has not yet been completed.
Right: Common behavior has been completed with now 32 X and 32 O. }
\end{figure}

\begin{figure}[t]
\begin{center}
\centerline{\epsfysize=7cm\epsfbox{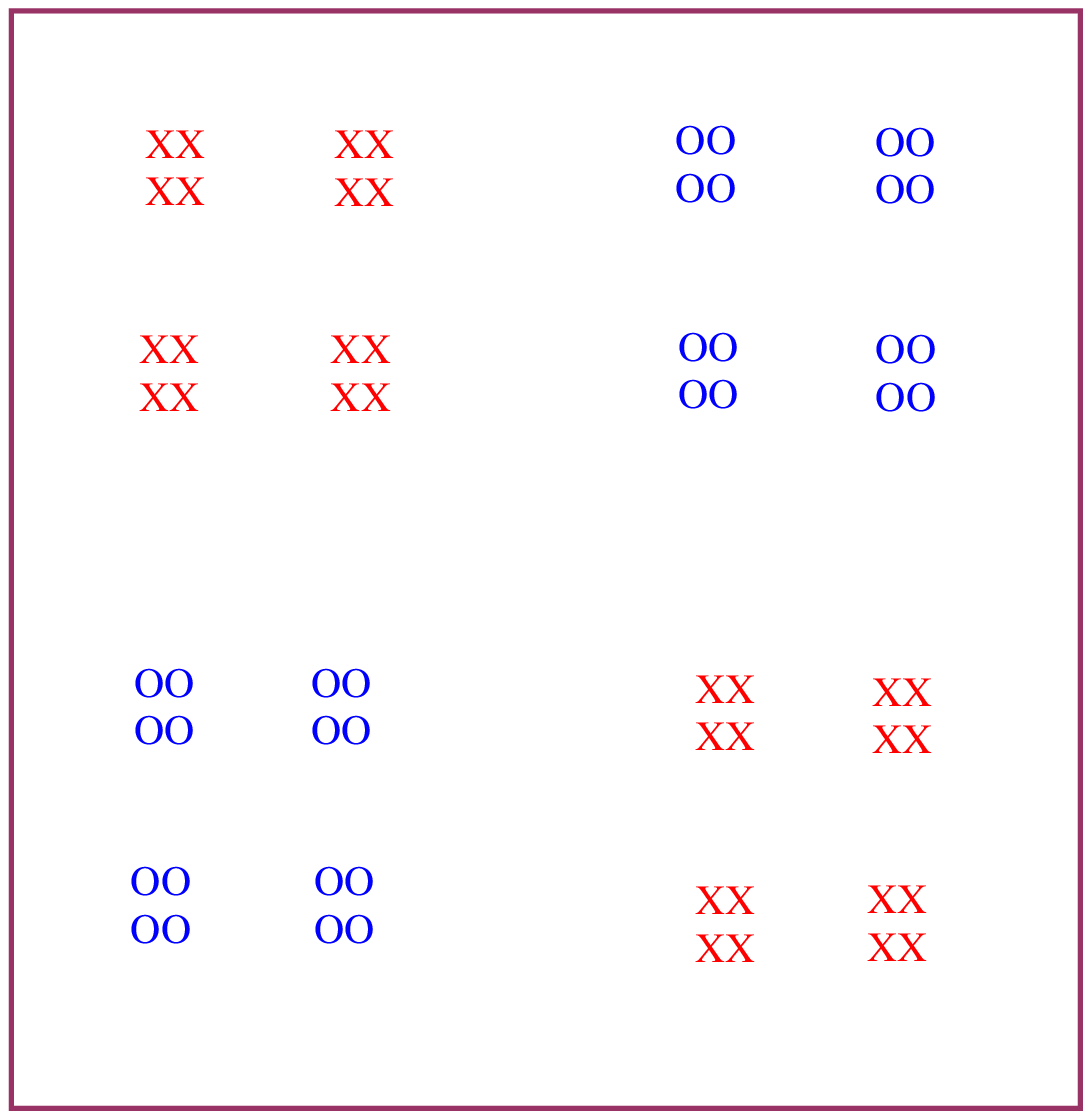}
\epsfysize=7cm\epsfbox{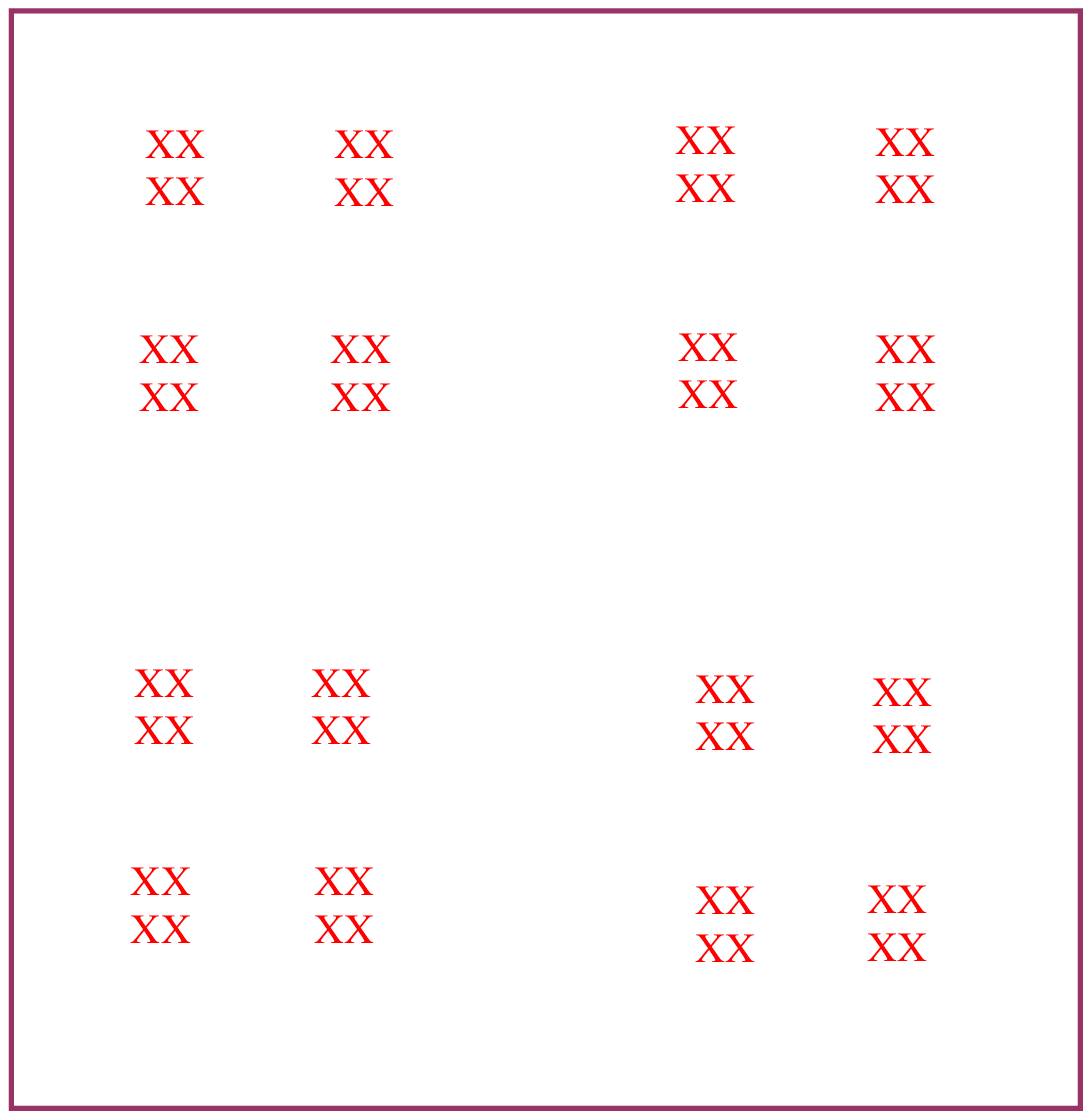}}
\end{center}
\caption{Left: Third and last level reference group is activated to include
the whole population of 64 persons. Common behavior has not yet been
completed. Right: Common behavior has been completed. The X wearing won. To
be compared with the right part of Figure (4) where the O wearing did win.}
\end{figure}

We can now see why it is important for a brand to multiply quality levels
and forms of marketing in its ''fili\`{e}re''. This strategy could be seen
as a classical process of differentiation. The wider the scale of products,
the larger the set of potential consumers. But each level can be considered
as a specific means of being seen by one or more members of a reference
group.

\subsection{ Organizing the ''fili\`{e}re''}

\begin{quote}
For a creator who wishes to invade the market, an investment threshold
exists, below which investing is useless, no matter what is the amount of
investment. Investing more is also superfluous. The optimal strategy
consists in hierarchically organizing the ''fili\`{e}re''.
\end{quote}

At this stage, the key issue is to determine the number of reference levels
needed to ensure full monopoly for the initial newcomer brand in the case of
small group sizes. For large size, the threshold is at around $50\%$ and the
level number is very small but with huge number of people involved. It is
also worth to notice that increasing the size to a larger group inceases the
threshold value toward that value of $50\%$. For instance we have $
p_{c}=0.35,0.40,0.42,0.44,0.45,0.46,0.46,0.47$ for respectively $
r=6,8,10,12,14,16,18,20$.

We can now calculate analytically the critical number of reference group
levels $n_{c}$ at which $p_{n_{c}}=\epsilon $ with $\epsilon $ being a very
small number. This determines the level of confidence of the prediction of
getting the whole market. One way to evaluate $n_{c}$ is to expand the
in-fashion function $p_{n}=P_{r}(p_{n-1})$ around the unstable fixed point $
p_{c}$, 
\begin{equation}
p_{n}\approx p_{c}+(p_{n-1}-p_{c})\lambda _{r}\ ,
\end{equation}
where $\lambda _{r}\equiv \frac{dP_{r}(p_{n})}{dp_{n}}|_{p_{c}}$ with $
P_{r}(p_{c})=p_{c}$. Rewriting the last equation as 
\begin{equation}
p_{n}-p_{c}\approx (p_{n-1}-p_{c})\lambda _{r}\ ,
\end{equation}
we can then repeat the process to get 
\begin{equation}
p_{n}-p_{c}\approx (p_{0}-p_{c})\lambda _{r}^{n}\ ,
\end{equation}
from which we get the critical number of levels $n_{c}$ at which $
p_{n}=\epsilon $. Taking the logarithm on both sides of Eq. (8) gives 
\begin{equation}
n_{c}\approx -\frac{\ln (p_{c}-p_{0})}{\ln \lambda _{r}}+n_{0}\ ,
\end{equation}
where $n_{0}\equiv \frac{\ln (p_{c}-\epsilon )}{\ln \lambda _{r}}$. As we
are dealing with integers, we have to take the integer part of Eq. (10).
Putting $n_{0}=1$ yields pretty accurate results.

\section{The effective optimal threshold}

As it may turn rather costly and complicated to modify both the size
reference groups and the associated number of levels we now evaluate the
following strategic issue. Given n reference group levels of group size r,
what is the initial overall proportion of B wearing required to get
everybody dressed in B?

It is worth keeping in mind the fact that the dynamics for A and B are not
symmetric. Here, we focus on the in-fashion dynamics from the B point of
view. To proceed, we invert Eq. (9) to obtain 
\begin{equation}
p_{0}=p_{c}+(p_{n}-p_{c})\lambda _{r}^{-n}\ .
\end{equation}
This yields two new operative critical thresholds. The first gives the value
of initial market penetration below which B is sure to disappear from the
market. It is obtained from Eq. (10) by inserting $p_{n}=0$ and it yields 
\begin{equation}
P_{D,r}^{n}=p_{c}(1-\lambda _{r}^{-n})\ .
\end{equation}

In parallel, setting $p_{n}=1$ gives the second threshold $P_{I,r}^{n}$,
above which the B brand invades the whole market. Combining Eqs. (11) and
(12) gives 
\begin{equation}
P_{I,r}^{n}=P_{D,r}^{n}+\lambda _{r}^{-n}\ .
\end{equation}
which shows the appearance of a strategic domain for $
P_{D,r}^{n}<p_{0}<P_{I,r}^{n}$. In this region B neither disappears totally
nor takes over the whole market. There is a coexistence with $p_{n}$ being
at equilibrium and neither 0 nor 1. This is therefore a coexistence region
in which both brands are present. No one is sure of winning. However, as
seen from Eq. (13), this coexistence region shrinks as a power law $\lambda
_{r}^{-n}$ of the number n of hierarchical reference group levels. Having a
small number of these levels raises the threshold for total invasion but
simultaneously lowers the threshold for disappearence.

The above formulas are approximated, since we have neglected corrections in
the vicinity of the stable fixed points. However they give the right
quantitative behavior. To get a practical feeling of what Eqs. (12) and (13)
mean in terms of numbers, let us illustrate them for the case $r=4$ where we
have $\lambda =1.64$ and $p_{c,4}=p_{c}=\frac{5-\sqrt{13}}{6}$. Considering
3, 4, 5, 6 and 7 level reference systems, $P_{D,r}^{n}$ is equals to
respectively $0.41$, $0.34$, $0.30$, $0.27$ and $0.26$. In parallel $
P_{I,r}^{n}$ equals $0.18$, $0.20$, $0.21$, $0.22$ and $0.22$. These series
dramatically emphasize the massive effect of the in-fashion process.

\section{Conclusion}

In this paper, we have discussed the different investment tools a creator
can use to invade a market, in other words to become the fashion. We use the
concept and techniques of real space normalisation group to study the
fashion diffusion process within hierarchical structures.We show that when
potential consumers are organised into large, anonymous groups producers
cannot recognize them. In this case, the level of investment making it
possible to invade the market is certainly high, and it may be optimal to
allow counterfeits. But allowing counterfeits can also be a dangerous tool.
We then show that if creators could organize people into distinctive
reference levels, this would diminish the threshold and make it possible to
avoid counterfeits. Simmel \cite{10} noted that the fashion market actually
seems to be the joint result of an imitation process and a search for
differentiation. We show that in creating different social leaders,
producers enable consumers to differentiate themselves from each other by
imitating their closest social leaders. To simplify our task, we have
focused on the clothes fashion market. With slight modifications, our
results could be extended to other kinds of fashion such as the fashion in
research ideas, for example.

\end{document}